\documentclass[prl,aps,showpacs,twocolumn]{revtex4}
\usepackage{graphicx}
\usepackage{dcolumn}
\usepackage{bm}
\usepackage{amsmath}
\usepackage{amssymb}
\usepackage{mathrsfs}
\usepackage{amsfonts}
\usepackage{url}

\begin{document} 

\title{Cavity QED of the graphene cyclotron transition}

\author{David Hagenm\"{u}ller}
\author{Cristiano Ciuti}
\affiliation{Laboratoire Mat\'eriaux et Ph\'enom\`enes Quantiques,
Universit\'e Paris Diderot-Paris 7 and CNRS, \\ B\^atiment Condorcet, 10 rue
Alice Domon et L\'eonie Duquet, 75205 Paris Cedex 13, France}

 \date{\today} 
\begin{abstract}
We investigate theoretically the cavity quantum electrodynamics of the cyclotron transition for Dirac fermions in graphene. We show that the ultrastrong coupling regime characterized by a vacuum Rabi frequency comparable or even larger than the transition frequency can be obtained for high enough filling factors of the graphene Landau levels. Important qualitative differences occur with respect to the corresponding physics of massive electrons in a semiconductor quantum well. In particular, an instability for the ground state analogous to the one occuring in the Dicke-model is predicted for increasing value of the electron density.
\end{abstract}

\maketitle

Cavity  Quantum Electrodynamics (QED) in the ultrastrong coupling regime is a fascinating topic that is attracting considerable interest in condensed matter physics, particularly in semiconductor microcavities\cite{Haroche,Ciuti,Anappara,Gunter,Todorov} and superconducting circuit QED systems\cite{Devoret,Niemczyk,Fedorov,Nataf1}. The ultrastrong coupling regime is achieved when the vacuum Rabi frequency (quantifying the interaction between one cavity photon and one elementary electronic excitation) becomes comparable or even larger than the corresponding electronic transition frequency. In such regime, it is possible to manipulate the quantum ground state of the cavity system, to modify the decoherence properties of the system\cite{Nataf2} and to enhance interesting non-adiabatic cavity QED effects\cite{DeLiberato,Gunter}. 

Recently, it was theoretically predicted that by coupling a cavity photon mode to the cyclotron transition (of frequency $\omega_0$) of a two-dimensional electron gas (2DEG) in a semiconductor, it is possible to have a vacuum Rabi frequency $\Omega_0$ such that $\Omega_0 / \omega_0 \sim \sqrt{\alpha_{\rm fs} \, \nu\, n_{\rm QW}}$ where $\alpha_{\rm fs}$ is the fine structure constant, $\nu$ is the filling factor of the Landau levels and $n_{\rm QW}$ is the (effective) number of quantum wells\cite{Hagenmuller}. Such a predicted scaling has been quantitatively demonstrated by recent impressive spectroscopy experimental results in the THz domain\cite{Scallari} .The striking consequence of this physical behavior is that for high filling factors it is possible to have $\Omega_0/\omega_0 \gg 1$.  In this ultrastrong coupling limit for the 2DEG, however, no instability is expected for the ground state, as a result of the role played by the so-called diamagnetic ${\bf \hat{A}^2}$-term\cite{Nataf3} (${\bf \hat{A}}$ is the electromagnetic vector potential operator). In the case of massive electrons in a semiconductor, the effective mass approximation is known to work very well in a broad range of conditions and one can generally consider just the conduction band to describe the physics of the two-dimensional electron gas, hence the underlying crystal structure of the semiconductor host turns out to be unimportant for many physical effects. In the case of graphene with massless fermions\cite{Wallace,Zhang,Novoselov}, this is certainly not the case. Some recent experimental works\cite{Zhang,Berger,Neugebauer} have demonstrated a relatively high carrier mobility in different graphene devices in the high density regime ($\gtrsim 10^{4} {\rm cm}^{2} {\rm V}^{-1} {\rm s}^{-1}$) leading a well defined cyclotron resonance even in the case of small magnetic fields. A clearly intriguing problem is to explore how graphene behaves when embedded in a cavity resonator. In particular, is it possible to achieve ultrastrong coupling between a cavity photon and the cyclotron transition in graphene ? If yes, is there any qualitative difference with respect to the case of massive fermions in semiconductors ?

In this letter, we present a microscopic theory for the cavity QED of graphene under perpendicular magnetic field, giving an answer to the questions formulated above. Through a  quantum field approach (accounting also for Coulomb depolarization effects) we show that the ultrastrong coupling regime is achievable for graphene. Moreover, due to the emergence of the Dirac cones with linear energy dispersions, the role of the ${\bf \hat{A}^2}$ turns out to be negligible when the lattice constant is much smaller than the photon wavelength. We predict the occurrence of a vacuum instability, which is absent in the case of the 2DEG of massive electrons.

Graphene is a single layer of carbon atoms forming a honeycomb lattice with two equivalent triangular Bravais sublattices $A$ and $B$. Each atom of $A$ ($B$) type is connected to its $3$ nearest neighbors of $B$ ($A$) type  via the displacement vectors ${\bm e}_{1} = a \sqrt{3}/2 {\bm u}_{x} + a/2 {\bm u}_{y}$, ${\bm e}_{2} = -a \sqrt{3}/2 {\bm u}_{x} + a/2 {\bm u}_{y}$ and ${\bm e}_{3} = -a {\bm u}_{y}$. We consider here the tight-binding description  taking only the first nearest neighbors into account with a hopping parameter $t$. In this case, the first quantization Hamiltonian in  ${\bm Q}$-space can be expressed in a two-components spinor basis due to the sublattice isospin\cite{Castro}, namely:  

\begin{equation}
\label{begin}
\mathcal{H}_{\bm Q} =\left( \begin{array}{cc}
0 & h_{\bm Q} \\ h^{*}_{\bm Q}  & 0 \end{array} \right) 
\end{equation}

where $h_{\bm Q}  = -t \sum^{3}_{j=1} e^{-i {\bm Q} \cdot {\bm e}_{j}}$. In the vicinity of the two inequivalent Dirac points (valleys) $K$ and $K'$ marked by the two vectors ${\bm K}^{\pm}=\pm4 \pi/(3\sqrt{3} a) {\bm u}_{x}$, the low-energy excitations are well described by the massless Dirac Hamiltonian $\mathcal{H}_{\alpha,{\bm \kappa}} = \hbar v_{F} \left(\alpha \kappa_x \sigma_x + \kappa_y \sigma_y \right)$ where ${\bm \kappa}={\bm Q}-{\bm K}^{\pm}$, $\sigma_{i}$ ($i=x,y,z$) are the Pauli matrices, $v_F=(3at)/(2\hbar)$ is the Fermi velocity ($\sim 10^6 {\rm m/s}$) and $\alpha=\pm$ is the valley isospin index. 

Here, we are interested in the case where a static magnetic field ${\bm B}$ is applied perpendicularly to the graphene plane. This problem can be exactly solved by means of the Peierls substitution (which remains valid as long as $a$ is much smaller than the magnetic length $l_{0}=\sqrt{(\hbar c)/(eB)}$ in the Dirac Hamiltonian), i.e. replacing ${\bm \kappa}$ with ${\bm \Pi}_{0}={\bm p} + \frac{e}{c} {\bm A}_{0}$ where ${\bm A}_{0}=-By{\bm u}_{x}$ is the vector potential (Landau gauge). This yields the graphene Landau Levels \cite{McClure}  whose energies are $E^{\lambda=\pm}_{n}= \lambda \hbar \omega_{0} \sqrt{n}$ where $n \geq 0$, $\lambda$ refers to the electron ($\lambda=+$) or hole band ($\lambda=-$) and $\omega_{0}=v_{F} \sqrt{2} / l_{0}$. Each Landau level has a degeneracy $\mathcal{N}=g_{s}S/(2 \pi l^{2}_{0})$, $S$ being the surface of the graphene layer and $g_{s}=4$ accounts for the spin and valley isospin degeneracy (we neglect Zeeman and valley splittings). Finally we define the filling factor (we use a different convention than the usual one\cite{Zhang,Novoselov}) as $\nu=\rho S/ \mathcal{N} + 1/2$ ($\rho$ is the total electron density) so that $\nu=n $ corresponds to the situation where the last fully occupied Landau level has an orbital quantum number $n-1$. In the following, we will study the case of integer filling factors and consider for simplicity the case of Fermi level in the electron band ($\lambda=+$). Moreover, we will consider a graphene layer embedded in a 0D cavity resonator. We investigate the coupling between a particular cavity photon mode and the transition between the last occupied Landau level $n=\nu-1$ and the first unoccupied one $n=\nu$ with a frequency $\omega_{0} \Delta_{\nu}=\omega_{0} \left(\sqrt{\nu}-\sqrt{\nu-1}\right)$. The Dirac fermion field operator in the valley $\alpha$ is written as a two-components spinor

\begin{equation}
\label{field_fermion}
{\bm  \hat{\Psi}}_{\alpha} ({\bm r}) = \sum_{\lambda,n,k} \left( \begin{array}{c} \psi^{(A)}_{\lambda,n,k,\alpha} ({\bm r}) \\ \\ \psi^{(B)}_{\lambda,n,k,\alpha} ({\bm r}) \end{array} \right) c_{\lambda,n,k,\alpha},
\end{equation}
where the wavefunctions in the valley $K$ are 
\begin{equation}
\label{valley1}
\left( \begin{array}{c} \psi^{(A)}_{\lambda,n,k,+} ({\bm r}) \\ \\ \psi^{(B)}_{\lambda,n,k,+} ({\bm r}) \end{array} \right) = \frac{e^{ikx}}{\sqrt{L}} \left( \begin{array}{c} - \lambda C^{-}_{n} \varphi_{n-1,k} (y)\\ \\ C^{+}_{n} \varphi_{n,k} (y) \end{array} \right) \delta \left( z - \frac{L_{z}}{2} \right).
\end{equation}

The operator $c_{\lambda,n,k,\alpha}$ annihilates a fermion in the state labelled by the quantum numbers $\left(\lambda,n,k,\alpha\right)$. $C^{-}_{n}=\sqrt{1-\delta_{n,0}/2}$ and $C^{+}_{n}=\sqrt{1+\delta_{n,0}/2}$ are the normalization factors, $L$ is the length of the graphene sheet (taken squared for simplicity) along the $x$ and $y$ directions, $L_{z}$ is the cavity length along the $z$ direction and $\varphi_{n,k} (y)$ corresponds to the eigenfunction of the 1D harmonic oscillator problem shifted by the guiding centre position $y_0 =k l^{2}_{0}$. The wavefunctions in the valley $K'$ are obtained by swapping the spinor (\ref{valley1}) and changing $\lambda \to -\lambda$.  
 
\begin{figure}[h!]
\label{sketch}
\begin{center}
\includegraphics[width=240pt]{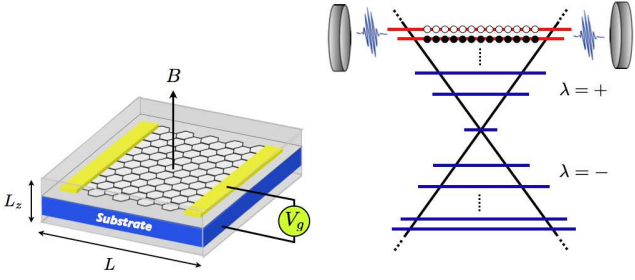}
\caption{\it Left panel: sketch of a cavity resonator of volume $V=L_z L^{2}$  ($L_z \ll L$) embedding a graphene layer with an uniform and static magnetic field $B$ perpendicular to the atomic plane. Right panel: a fully confined cavity photon mode is supposed to be quasi-resonant to the graphene cyclotron transition between LLs $n=\nu-1$ and $n=\nu$ (red horizontal solid lines). \label{cavity}}
\end{center}
\end{figure}

We consider a 0D cavity resonator (see Fig. \ref{cavity}) of volume $V=L_zL^{2}$ that confines the electromagnetic modes along the three spatial directions. These cavity modes are labelled by the quantized wavevector ${\bm q}\equiv \left(q_{x},q_{y},q_{z}\right) \equiv \left(\pi n_{x}/L,\pi n_{y}/L,\pi n_{z}/L_{z}\right)$, where $n_x$, $n_y$, $n_z$ are integer numbers. For sake of simplicity, we will consider that the cavity length $L_z$ along the $z$ direction is much smaller than the cavity transverse size $L$. This way, we can restrict our study to the particular photon mode with $n_z=1$, neglecting all the higher-lying modes ($n_z > 1$). We will also assume that the graphene layer is placed in the middle of the cavity at $z=L_{z}/2$ (see Fig. \ref{cavity}). Of course, our theory can be easily generalized to a more complicated geometry. The electromagnetic vector potential reads \cite{Kakazu}

\begin{equation}
\label{field_photon}
{\bf \hat{A}}_{em}({\bm r}) =\sum_{\bm{q},\eta} \sqrt{\frac{2\pi\hbar c^{2}}{\epsilon \, \omega_{\rm cav} ({\bm q}) V}} \left(a_{\bm{q},\eta} {\bm u}_{\bm{q},\eta}+ a^{\dagger}_{\bm{q},\eta} {\bm u}^{*}_{\bm{q},\eta} \right)
\end{equation}
where $\eta=1,2$ is the photon polarization, $\omega_{\rm cav} ({\bm q})$ is the cavity frequency, $\epsilon$  the cavity dielectric constant and \cite{note_mode}
 
\begin{equation}
\label{optical_mode_1}
{\bm u}_{{\bm q},1}  =  \left(
\begin{array}{c}
C_{q_x} \cos (q_{x} x) \sin (q_{y} y) \sin (\pi z/L_z) \, \cos \theta \cos \phi \\ C_{q_y} \sin (q_{x} x) \cos (q_{y} y) \sin (\pi z/L_z) \, \cos \theta \sin \phi \\ - 2 \sqrt{2} \sin (q_{x} x) \sin (q_{y} y) \cos (\pi z/L_z) \, \sin \theta \end{array} \right), 
\end{equation}
\begin{equation}
\label{optical_mode_2}
{\bm u}_{{\bm q},2}  =  \left( 
\begin{array}{c}
- C_{q_x} \cos (q_{x} x) \sin (q_{y} y) \sin (\pi z/L_z) \, \sin \phi \\ C_{q_y} \sin (q_{x} x) \cos (q_{y} y) \sin (\pi z/L_z) \, \cos \phi \\ 0 \end{array} \right),
\end{equation}

with $\cos \theta = q_z/\vert {\bm q} \vert$ and $\cos \phi = q_x/\sqrt{q_x^2+q_y^2}$.
The operator $a_{\bm{q},\eta}$ is the photon annihilation operator in the cavity mode $\left({\bm q},\eta \right)$.  

When considering the graphene sheet in the presence of a perpendicular static magnetic field embedded in our cavity resonator, the electron velocity operator involves the total vector potential ${\bf \hat{A}}_{\rm T} = {\bf A}_{0} + {\bf \hat{A}}_{em}$. The coupling Hamiltonian is obtained by replacing the momentum ${\bm p}$ in the free Hamiltonian by its gauge invariant form ${\bm \Pi}= {\bm p} + \frac{e}{c} {\bm A}_{\rm T}={\bm \Pi}_{0} + \frac{e}{c} {\bm A}_{\rm em}$. The Hamiltonian (\ref{begin}) becomes 

\begin{equation}
\label{place}
h_{\alpha} =-t\sum_{j=1}^{3} Z^{(\alpha)}_{j} \exp{\left[-\frac{i}{\hbar} \left({\bm \Pi}_{0} \cdot {\bm e}_{j} + \frac{e}{c} {\bm A}_{\rm em} \cdot {\bm e}_{j} \right)\right]},
\end{equation}

where we have defined the valley-dependent phase factor $Z^{(\alpha)}_{j}=e^{- i {\bm K}^{\pm}\cdot {\bm e}_{j}}$. As we are interested in the quantum ground state properties and low-energy excitations, we can consider the continuum limit (lattice size $a \to 0$, which is the relevant limit when $a$ is much smaller than any other length scale). In other words, we expand the Hamiltonian (\ref{place}) with respect to the lattice parameter $a$ and retain the leading contribution. This way, we get two terms, the first one is the massless Dirac Hamiltonian $\mathcal{H}^{(\alpha)}_{\rm 0}$ with static magnetic field, while the second one $\mathcal{H}^{(\alpha)}_{\rm int}$ represents the coupling between the Dirac fermions and the cavity optical modes:

\begin{eqnarray}
\label{free}
\mathcal{H}^{(\alpha)}_{\rm 0} &=& \hbar v_F \left( \alpha \Pi_{{\rm 0},x} \sigma_{x} + \Pi_{{\rm 0},y} \sigma_{y} \right) \\ 
\label{int}
\mathcal{H}^{(\alpha)}_{\rm int}&=&\frac{\hbar v_{F}e}{c} \left( \alpha A_{{\rm em},x} \sigma_{x} + A_{{\rm em},y} \sigma_{y} \right)
\end{eqnarray}

Note that the coupling Hamiltonian (\ref{int}) depends linearly on the bosonic operators $a_{\bm{q},\eta}$ and $a^{\dagger}_{\bm{q},\eta}$. Consequently, the total Hamiltonian does not contain any term involving the squared vector potential ${\bm A}^{2}_{\rm em}$ as it is usually the case in cavity QED for massive electrons. Using Eqs. (\ref{field_fermion}), (\ref{field_photon}) and (\ref{int}), the second quantized coupling Hamiltonian 
$
H_{\rm int}=\int \!\! d^{2} r \, {\bf  \hat{\Psi}}^{\dagger} ({\bm r}) \mathcal{H}^{(\alpha)}_{\rm int} {\bf \hat{\Psi}} ({\bm r})
$ becomes valley independent so we can omit the index $\alpha$ taking the valley degeneracy into account through the factor $g_{s}$. Note that since we are dealing with optical modes, we have $\vert {\bm q} \vert l_{0} \ll 1$. This condition allows us to neglect the LL mixing\cite{Roldan1} when considering only the transition $\nu \to \nu-1$. The final result is

\begin{equation}
\label{coupling}
H_{\rm int} = \sum_{{\bm q},\eta} \hbar \Omega_{{\bm q},\eta}  \left(a_{{\bm q},\eta} + a^{\dagger}_{{\bm q},\eta} \right) \left(d_{{\bm q},\eta} + d^{\dagger}_{{\bm q},\eta} \right) 
\end{equation}

where $\Omega_{{\bm q},1}=-\Omega_{\bm q} \cos \theta$, $\Omega_{{\bm q},2} = \Omega_{\bm q}$ with the vacuum Rabi frequency (for $\nu > 1$) given by

\begin{equation}
\label{coupling_0}
\Omega_{\bm q} = \omega_0 \sqrt{\frac{ \alpha_{\rm fs}  g_{s} }{4 \pi \sqrt{\epsilon}}}.
\end{equation}

The boson annihilation operators $d_{{\bm q},\eta}$  (polarization $\eta=1,2$) corresponding to the bright collective modes are:
\begin{widetext}
\begin{eqnarray}
\label{trans_1}
d_{{\bm q},1} & = & \frac{1}{\sqrt{\mathcal{N}}} \sum_{k} \sin \left[ q_{y} \left(k+\frac{q_x}{2} \right)l^{2}_{0} - \phi  \right] c^{\dagger}_{\nu-1,k} c_{\nu,k+q_{x}} + \frac{1}{\sqrt{\mathcal{N}}} \sum_{k} \sin \left[ q_{y} \left(k-\frac{q_x}{2} \right)l^{2}_{0} + \phi \right] c^{\dagger}_{\nu-1,k} c_{\nu,k-q_{x}},  \\ 
\label{trans_2}
d_{{\bm q},2}  &=& \frac{1}{\sqrt{\mathcal{N}}} \sum_{k} \cos \left[ q_{y} \left(k+\frac{q_x}{2} \right)l^{2}_{0} - \phi  \right] c^{\dagger}_{\nu-1,k} c_{\nu,k+q_{x}} - \frac{1}{\sqrt{\mathcal{N}}} \sum_{k} \cos \left[ q_{y} \left(k-\frac{q_x}{2} \right)l^{2}_{0} + \phi \right] c^{\dagger}_{\nu-1,k} c_{\nu,k-q_{x}} .
\end{eqnarray}
\end{widetext}

The first relevant conclusion of our letter is that graphene can enter deeply the ultrastrong coupling regime. In fact, in the high filling factor regime (i.e. $\nu \gg 1$) the cyclotron transition frequency is $\omega_0 \Delta_{\nu} \sim \omega_0/(2 \sqrt{\nu})$. Hence, the vacuum Rabi frequency normalized to the transition frequency 

\begin{equation}
\frac{\Omega_{\bm q}}{\omega_0 \Delta_{\nu}}\simeq \sqrt{\nu} \sqrt{\frac{ \alpha_{\rm fs} g_{s}}{ \pi \sqrt{\epsilon}}} \end{equation}

can be larger than $1$ for large enough filling factors. To complete our treatment, we consider the Coulomb interaction following a bosonization procedure\cite{Westfahl,Doretto,Roldan2} allowing us to calculate the depolarization shift (Random phase approximation (RPA) contribution). The second quantized Coulomb Hamiltonian reads:

\begin{equation}
\label{coulomb}
H_{\rm coul} = \frac{1}{2} \int \!\!\! \int \!\! d^{2} r \, d^{2} r' \, {\bf  \hat{\Psi}}^{\dagger} ({\bm r}) {\bf  \hat{\Psi}} ({\bm r}) \frac{e^{2}}{\epsilon \vert {\bm r}-{\bm r'} \vert} {\bf  \hat{\Psi}}^{\dagger} ({\bm r'}) {\bf  \hat{\Psi}} ({\bm r'}). 
\end{equation}

In the considered geometry, it is convenient to expand the Coulomb potential $V({\bm r}-{\bm r'})$ using a 2D Fourier series. Finally, we use the same procedure as for the light-matter coupling Hamiltonian derivation restricting the Coulomb Hamiltonian to describe only scattering processes between LLs $\nu-1$ and $\nu$ thanks to the condition $\vert {\bm q} \vert l_{0} \ll 1$. This provides the Coulomb Hamiltonian written in terms of the collective modes $d_{{\bm q},\eta}$ and $e_{{\bm q},\eta}$ 

\begin{equation}
\label{cool}
H_{\rm coul} = \sum_{{\bm q},\eta} \hbar V_{\bm q} \gamma_{\eta} \left[\left(d^{\dagger}_{{\bm q},\eta}+\gamma_{\eta}d_{{\bm q},\eta}\right)^{2} - \left(e^{\dagger}_{{\bm q},\eta}-\gamma_{\eta}e_{{\bm q},\eta}\right)^{2} \right], 
\end{equation}
with $\gamma_{1} = -1$, $\gamma_{2} = 1$ and
where $V_{\bm q}= \frac{\alpha_{\rm fs} g_{s} c \vert {\bm q}_{\perp} \vert}{32 \epsilon}\xi^{2}_{\nu}$, ${\bm q}_{\perp}=\left(q_{x}, q_{y} \right)$ denotes the in-plane wavevector and $\xi_{\nu} = C^{-}_{\nu-1}\sqrt{\nu-1} + C^{+}_{\nu-1}\sqrt{\nu}$. In the high filling factor regime, we have $\xi_{\nu} \sim \sqrt{2\nu}$.

The Coulomb Hamiltonian reveals the presence of two additional dark modes (i.e uncoupled to the electromagnetic field) corresponding to two operators $e_{{\bm q},\eta}$ and $e^{\dagger}_{{\bm q},\eta}$. Since the bright and dark modes are decoupled by the Coulomb interaction, if we look at the ground state properties, we can omit the dark modes. After bosonizing the kinetic energy\cite{Westfahl,Doretto}, the Hamiltonian reads:

\begin{eqnarray}
H_{\rm dp}&=&\sum_{{\bm q},\eta} \hbar \omega_{0} \Delta_{\nu} d^{\dagger}_{{\bm q},\eta} d_{{\bm q},\eta} + \hbar V_{\bm q} \gamma_{\eta} \left(d^{\dagger}_{{\bm q},\eta}+\gamma_{\eta} d_{{\bm q},\eta}\right)^{2} \nonumber \\
\label{plasmon}
&=& \sum_{{\bm q},\eta} \hbar \omega_{p} ({\bm q}) \; m^{\dagger}_{{\bm q},\eta} m_{{\bm q},\eta} + \text{const}.
\end{eqnarray}

where $m_{{\bm q},1}$ and $m_{{\bm q},2}$ corresponds to the magnetoplasmon modes with energy $\omega_{p} ({\bm q}) =\sqrt{\omega_{0} \Delta_{\nu} \left(\omega_{0} \Delta_{\nu} + 4 V_{\bm q}\right)}$. Since our bosonic approach allows us to calculate the RPA contribution\cite{Westfahl}, we find that $V_{\bm q} \to 0$ when ${\bm q} \to 0$. However, our treatment does not include the vertex corrections (attractive  electron-hole Coulomb interaction ) nor the difference between the exchange self-energy of the electron and that of the hole. For this reason, we do not find the renormalization of the transition frequency $\omega_{0} \Delta_{\nu}$ at zero wavevector which is expected since the Kohn's theorem\cite{Kohn} does not apply in graphene\cite{Roldan1}. Nevertheless, some recent theoretical studies\cite{Iyengar,Bychkov,Gorbar} have shown that these corrections lead to a relatively moderate renormalization ($< 20\%$) of the Fermi velocity and can thus be easily accounted for in our model. In terms of the magnetoplasmon modes, the Hamiltonian reads:
\begin{eqnarray}
\label{total}
H_{\rm tot}&=&\sum_{{\bm q},\eta} \hbar \omega_{p} ({\bm q}) \; m^{\dagger}_{{\bm q},\eta} m_{{\bm q},\eta} + \hbar \omega_{\rm cav} ({\bm q}) a^{\dagger}_{{\bm q},\eta} a_{{\bm q},\eta} \nonumber \\ 
&+& \hbar \Lambda_{{\bm q},\eta} \left(a_{{\bm q},\eta} + a^{\dagger}_{{\bm q},\eta} \right)\left( m_{{\bm q},\eta} + m^{\dagger}_{{\bm q},\eta} \right)
\end{eqnarray}
where $\Lambda_{{\bm q},1} = \Omega_{\bm q}\cos \theta \sqrt{\frac{\omega_{p} ({\bm q})}{\omega_{0} \Delta_{\nu}}}$ and $\Lambda_{{\bm q},2} = \Omega_{\bm q} \sqrt{\frac{\omega_{0} \Delta_{\nu}}{\omega_{p} ({\bm q})}}$.

In contrast to the Hopfield\cite{Hopfield} quantum Hamiltonian for massive quasi-particles in semiconductor microcavities\cite{Ciuti,Todorov}, such an Hamiltonian is reminiscent of the Dicke Hamiltonian, which admits a quantum critical point (QCP) beyond which the normal ground state becomes unstable. Diagonalizing the Hamiltonian (\ref{total}) via the Bogoliubov method, we find that the energy of the lower eigenvalue vanishes (existence of a gapless excitation) when $\Omega_{\bm q}=\sqrt{\frac{\omega_{0} \Delta_{\nu}\omega_{\rm cav}({\bm q})}{4}}$ for the branch $\eta=1$ and $\Omega_{\bm q}=\sqrt{\frac{\omega^{2}_{p} ({\bm q}) \omega_{\rm cav}({\bm q})}{4\omega_{0} \Delta_{\nu}}}$ for the branch $\eta=2$. Note that the no-go theorem for cavity QED quantum phase transitions\cite{Nataf3} holds as long as the effective kinetic energy takes the usual quadratic form ${\bm p}^{2}/2m^{*}$ which well describes massive electrons in semiconductors ($m^{*}$ is the electron band effective mass). In graphene, the peculiar crystal structure is essential in determining the electronic properties and the effective kinetic energy is governed by the massless Dirac Hamiltonian\cite{Wallace,Castro} $v_{F} {\bm p}\cdot {\bm \sigma}$ at low energy (${\bm \sigma}=\left( \sigma_{x},\sigma_{y}\right)$ being the Pauli matrices). Since this Hamiltonian is linear in ${\bm p}$, such no-go theorem does not necessarily apply and indeed our work (which starts from the tight-binding Hamiltonian of graphene) predicts that a vacuum instability occurs. 

\begin{figure}[h!]
\begin{center}
\includegraphics[width=240pt]{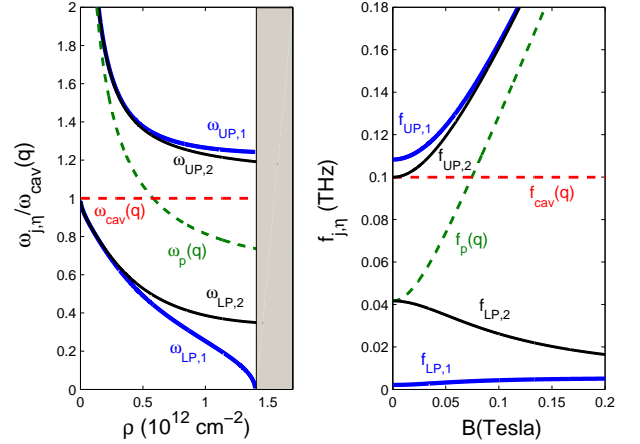}
\caption{\label{plot} Left panel: normalized frequencies $\omega_{j,\eta}/\omega_{\rm cav} ({\bm q})$ (solid lines) of the graphene magnetopolariton lower and upper branches as a function of the carrier density for a given magnetic field $B=50 \, {\rm mT}$. The dashed lines depict the bare cavity and magnetoplasmon frequencies. Parameters:   $\epsilon=3.9$, $L_{z}=760 \, {\rm \mu m}$ and $L=5L_{z}$, optical mode with ${\bm q}\equiv \left(n_{x}=1,n_{y}=1,n_{z}=1\right)$ and  $\omega_{\bm q}=0.628 \, {\rm THz} \, {\rm rad}^{-1}$ ($f_{\bm q}=0.1 \, {\rm THz}$). Right panel: same quantities (unnormalized and divided by $2\pi$) as a function of the magnetic field just below the critical density $\rho_{c}=1.4 \times 10^{12} {\rm cm}^{-2}$. }
\end{center}
\end{figure}

The left panel of the Fig. \ref{plot} shows the frequencies of the magnetopolariton excitations $\omega_{j,\eta}$, ($j=LP,UP$) normalized to the cavity mode frequency versus electron density. We see that by increasing the density, the lower branch $\eta=1$ has a nonmonotonical behavior and vanishes when reaching a critical value\cite{note_rho} $\rho_{c}=1.4 \times 10^{12} {\rm cm}^{-2}$. On the right panel, the magnetopolariton frequencies $f_{j,\eta}=\omega_{j,\eta}/2\pi$ are depicted as a function of the magnetic field just below the critical density $\rho_{c}$. We notice a strongly asymmetric dispersion when increasing the magnetic field together with the vanishing frequency of the lower branch $f_{\rm LP,1}$. These characteristics are a signature of such a vacuum instability. In order to describe the properties of the Dicke Hamiltonian above the QCP, an Holstein-Primakoff\cite{Holstein} bosonization procedure\cite{Emary} can be applied. However, such a mapping does not hold in this case because of the non-vertical excitations at finite wavevector\cite{Doretto}, making the analytical description of the excitations above the QCP highly non-trivial and will be the scope of future investigations. 

In conclusion, we have shown that the peculiar crystal structure of graphene leads to remarkable cavity QED properties. In particular, the coupling of the graphene cyclotron transition is qualitatively different for the case of massive electrons in semiconductors. In the ultrastrong coupling regime, a vacuum instability analogous to the one occuring in the Dicke-model can also occur for graphene, whereas for massive quasi-particles this is not the case. In free space graphene (without cavity), the anomalous properties of Dirac quasiparticles have been shown to lead to an unusual behavior of the optical conductivities in the presence of a perpendicular magnetic field\cite{Gusynin}. For cavity embedded Dirac fermions, our work paves the way to  interesting developments, such as cavity-controlled magnetotransport in graphene. We would like to thank M. O. Goerbig, S. De Liberato and P. Nataf for fruitful discussions.


\begin{thebibliography}{0}
\expandafter\ifx\csname natexlab\endcsname\relax\def\natexlab#1{#1}\fi
\expandafter\ifx\csname bibnamefont\endcsname\relax
  \def\bibnamefont#1{#1}\fi
\expandafter\ifx\csname bibfnamefont\endcsname\relax
  \def\bibfnamefont#1{#1}\fi
\expandafter\ifx\csname citenamefont\endcsname\relax
  \def\citenamefont#1{#1}\fi
\expandafter\ifx\csname url\endcsname\relax
  \def\url#1{\texttt{#1}}\fi
\expandafter\ifx\csname urlprefix\endcsname\relax\def\urlprefix{URL }\fi
\providecommand{\bibinfo}[2]{#2}
\providecommand{\eprint}[2][]{\url{#2}}

\end{thebibliography}


\begin{thebibliography}{99}

\bibitem{Haroche}
S. Haroche and J.-M. Raimond, \textit{Exploring the quantum: atoms, cavities, photons}, (Oxford Press, 2006).

\bibitem{Ciuti}
C. Ciuti, G. Bastard, and I. Carusotto, Phys. Rev. B {\bf 72}, 115303 (2005).

\bibitem{Anappara}
A. A. Anappara \textit{et al.}, Phys. Rev. B {\bf 79}, 201303(R) (2009).

\bibitem{Gunter}
G. G\"{u}Ÿnter \textit{et al.}, Nature (London) {\bf 458}, 178 (2009).

\bibitem{Todorov}
Y. Todorov \textit{et al.}, Phys. Rev. Lett. {\bf 105}, 196402 (2010).

\bibitem{Devoret}
M. Devoret, S. Girvin, and Schoelkopf, Ann. Phys. {\bf 16}, 767 (2007).

\bibitem{Niemczyk}
T. Niemczyk \textit{et al.}, Nature Phys. {\bf 6}, 772-776 (2010).

\bibitem{Fedorov}
A. Fedorov \textit{et al.}, Phys. Rev. Lett. {\bf 105}, 060503 (2010).

\bibitem{Nataf1}
P. Nataf and C. Ciuti, Phys. Rev. Lett. {\bf 104}, 023601 (2010).

\bibitem{Nataf2}
P. Nataf and C. Ciuti, Phys. Rev. Lett. {\bf 107}, 190402 (2011).

\bibitem{DeLiberato}
S. De Liberato, C. Ciuti, and I. Carusotto, Phys. Rev. Lett. {\bf 98}, 103602 (2007).

\bibitem{Hagenmuller}
D. Hagenm\"{u}ller, S. De Liberato, and C. Ciuti, Phys. Rev. B {\bf 81}, 235303 (2010).

\bibitem{Scallari}
G. Scalari \textit{et al.}, \textit{submitted}; preprint  arXiv:1111.2486.

\bibitem{Nataf3}
P. Nataf and C. Ciuti, Nat. Commun. 1:72 (2010).

\bibitem{Wallace} 
P.R. Wallace, Phys. Rev. {\bf 71}, 622 (1947). 

\bibitem{Zhang}
Y. Zhang \textit{et al.}, Nature {\bf 438}, 201-204 (2005).

\bibitem{Novoselov}
K. S. Novoselov \textit{et al.}, Science {\bf 315}, 1379 (2007).

\bibitem{Berger}
C. Berger \textit{et al.}, Science {\bf 312}, 5777 (2006).

\bibitem{Neugebauer}
P. Neugebauer \textit{et al.}, Phys. Rev. Lett. {\bf 103}, 136403 (2009).

\bibitem{Castro}
A. H. Castro Neto \textit{et al.}, Rev. Mod. Phys. {\bf 81}, 109 (2009)

\bibitem{McClure}
J. W. McClure, Phys. Rev. {\bf 104}, 666 (1956).

\bibitem{Kakazu}
K. Kakazu and Y. S. Kim, Phys. Rev. A {\bf 50}, 1830 (1994).

\bibitem{note_mode}
The mode normalization is given by $C_{q_i}=2$ if $n_{i} = 0$ and $C_{q_i}=2 \sqrt{2}$ else.

\bibitem{Roldan1}
R. Rold\`{a}n, J.-N. Fuchs, and M.O. Goerbig, Phys. Rev. B {\bf 82}, 205418 (2010).

\bibitem{Westfahl}
H. Westfahl Jr., A. H. Castro Neto, and A. O. Caldeira, Phys. Rev. B {\bf 55}, R7347 (1997).

\bibitem{Doretto}
R. L. Doretto, A. O. Caldeira, and S. M. Girvin, Phys. Rev. B {\bf 71}, 045339 (2005).

\bibitem{Roldan2}
R. Rold\`{a}n, M.O. Goerbig, and J.-N. Fuchs, Phys. Rev. B {\bf 83}, 205406 (2011).

\bibitem{Kohn}
W. Kohn, Phys. Rev. {\bf 123}, 1242 (1961).

\bibitem{Iyengar}
A. Iyengar \textit{et al.}, Phys. Rev. B {\bf 75}, 125430 (2007).

\bibitem{Bychkov}
Yu. A. Bychkov and G. Martinez, Phys. Rev. B {\bf 77}, 125417 (2008).

\bibitem{Gorbar}
E. V. Gorbar \textit{et al.}, arXiv: 1105.1360v1.

\bibitem{Hopfield}
J. J. Hopfield, Phys. Rev. 112, 1555 (1958).

\bibitem{note_rho}
Considering the high filling factor limit ($\nu \gg 1$), the critical density is evaluated as $\rho_{c}=\left(\frac{\pi \sqrt{\pi \epsilon} \omega_{\rm cav}({\bm q})}{4 \alpha_{\rm fs}  v_{F} \sqrt{g_{s}}}\right)^{2}$. 

\bibitem{Holstein}
T. Holstein and H. Primakoff, Phys. Rev. {\bf 58}, 1098-1113 (1940).

\bibitem{Emary}
C. Emary and T. Brandes, Phys. Rev. E {\bf 67}, 066203 (2003).

\bibitem{Gusynin}
V. P. Gusynin and S. G. Sharapov, Phys. Rev. B {\bf 73}, 245411 (2006).

\end{thebibliography}
\end{document}